# Building an Entrepreneurship Data Warehouse

Yngve Dahle
Norwegian University of Science and Technology (NTNU)
Faculty of Engineering Science and Technology (IVT)
Norway
yngve.dahle@ntnu.no

Martin Steinert
Norwegian University of Science and Technology (NTNU)
Faculty of Engineering Science and Technology (IVT)
Norway
martin.steinert@ntnu.no

Anh Nguyen Duc
Norwegian University of Science and Technology (NTNU)
Faculty of Information Technology and Electronic (IME)
Norway
anhn@ntnu.no

Pekka Abrahamsson
Norwegian University of Science and Technology (NTNU)
Faculty of Information Technology and Electronic (IME)
Norway
pekkaa@ntnu.no

*Abstract*—The main principle of the Lean Startup movement is that static business planning should be replaced by a dynamic development, where products, services, business model elements, business objectives and activities are frequently changed based on constant customer feedback. Our ambition is to empirically measure if such changes of the business idea, the business model elements, the project management and close interaction with customers really increases the success rate of entrepreneurs, and in what way. Our first paper; "Does Lean Startup really work? - Foundation for an empirical study" presented the first attempt to model the relations we want to measure. This paper will focus on how to build and set up a test harness (from now on called the Entrepreneurship Platform or EP) to gather empirical data from Companies and how to store these data together with demographical and financial data from the PROFF-portal in the Entrepreneurial Data Warehouse (from now called the EDW). We will end the paper by discussing the potential methodological problems with our method, before we document a test run of our set-up to verify that we are actually able to populate the Data Warehouse with time series data.

*Keywords*: Big Data, Business Idea, Business Modelling, Data Warehouse, Machine Learning, Lean Startup, Pivot

## I. INTRODUCTION

Since 2011 a new paradigm within entrepreneurship has been established to challenge traditional waterfall models [16]. It is called Lean Startup, and is, among others, based on the writings of Eric Ries [13], Steve Blank [2], Alexander Osterwalder and Yves Pigneur [12] and Ash Maurya [9]. This paradigm differs from traditional business development in many ways, but the core of the concept can be interpreted as follows: For a startup, the rate of success correlates positively with:
1. Its ability to change the *business idea* when necessary.
2. Its ability to continuously change elements of its *business model* as defined by Osterwalder, Pigneur and Maurya. These two first points constitute a "pivot" as Eric Ries calls it.
3. Its ability to change elements of its *project development* according to changing circumstances
4. The frequency of its *customer interaction and testing* in the business development process.

In our previous work [6] the success of a startup was defined as a function of business idea, business model, project development and customer interaction:

$$S=\Delta f(BI)+\Delta f(BM) +\Delta f(PD)+\Delta f(CI)$$

Where S= Success, BI= Business Idea, BM= Business model, PD= Project Development and CI= Customer Interaction. This means that any alteration to the Business Idea, Business Model, Project Development and Customer Interaction will improve the level of Success.

We also outlined a possible way to build a test harness for gathering behavioural data from a large number of Companies over time (as shown in Figure 1). Utilizing a beta version of this tool, we did an initial pilot data gathering from a handful of Companies. This test verified that there are Companies that actually do change their Business Idea, Business Model and Project over time in a way similar to what is suggested in Lean Startup theory. This means it should be possible to evaluate the correlation between this behaviour and success.

This paper will focus on presenting how to build and set up the test harness, or the Entrepreneurial Platform (EP) in an optimal way and how to build the so-called Entrepreneurial Data Warehouse (EDW). We are encouraging the entrepreneurs to evaluate and to change the Business Idea, Business Model and Project Development as often as necessary, and to perform new Customer Interactions as often as they want. This design is based on the principles of the "wayfaring" theories described by [17].

Further we will show how we can populate the EDW by extracting and transferring data from the EP, and combining them with demographical and financial data from the PROFF-portal. We will end the paper by discussing the potential methodological problems we may encounter, before we document a test run of our set-up to verify that we are actually able to populate the Data Warehouse.

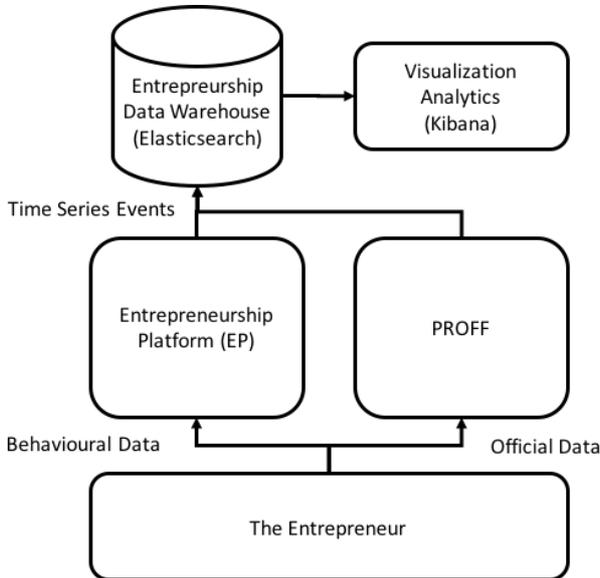

*Figure 1: Empirical data collection and analysis approach*

The paper is organized as follows, Section II presents the design of the Entrepreneurial Platform and the integration with PROFF. Section III presents the design of the Entrepreneurial Data Warehouse. Section IV discusses the challenges we may experience. Section V and VI offers a very brief initial run through of the first data entering the DW and describes some suggested future research.

II. THE ENTREPRENEURSHIP PLATFORM AND PROFF

A. *The Entrepreneurship Platform*

What set our project apart, is that we are creating a uniform categorization of the different theoretical events that constitutes a business development process. We call these *Event Categories*. The Entrepreneurship Platform (as shown in Figure 2) is the tool that enable us to monitor the creation of Events created, updated and deleted within each Event Category by a large number of entrepreneurs. After trying out different test versions in 2014, 2015 and 2016, the EP was released in its 1.0 version in January 2017 and in its Data Warehouse ready 1.1 version in late May 2017.

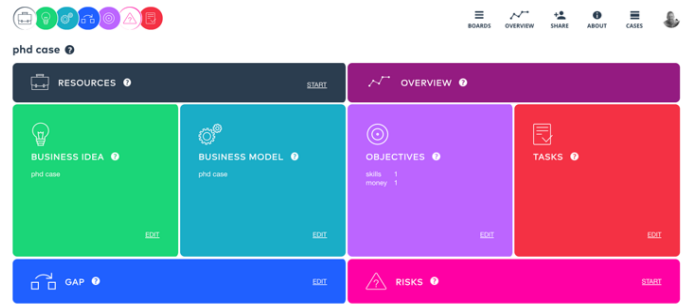

*Figure 2: The Entrepreneurial Platform*

The EP consists of approximately 150 components that constitutes 7 Boards, 3 different Customer Test Features and an Admin Console. Each of these consists of a number of *Boxes*, and the User Interface is based on filling these boxes with *Cards* containing the relevant Content. Everything is linked together, resulting in an *Overview*, giving the Case Company control over their current strategy, project and forecast. The thinking behind this model is clarified in the book Lean Business Planning [5].

The EP will be marketed directly as a free-of-cost SaaS tool toward the startups. In addition, it will be marketed in white-labelled versions from a number of partners (Banks, Innovation Centres, Venture Capitalists and Incubators). We are marketing a full version, a strategy only version and a project only version of the EP. The full version contains all the Boards, whereas the two other versions contains only the Boards relevant for the purpose. A Case Company can combine any of the seven Boards as they see fit. These different versions are called *Templates*. The Cases can switch between Templates. All variations will be flagged.

The Case Company can choose to work with a three-month. six-month or a 12-month fixed or rolling Case period. They can also invite internal and external participants to co-work with the Case. The internal participants are typically colleagues. The external participants can be representatives from helpers such as banks, incubators, consultants or innovation centres. We want to catch this whole level of interaction between the whole entrepreneurial ecosystem in our data set.

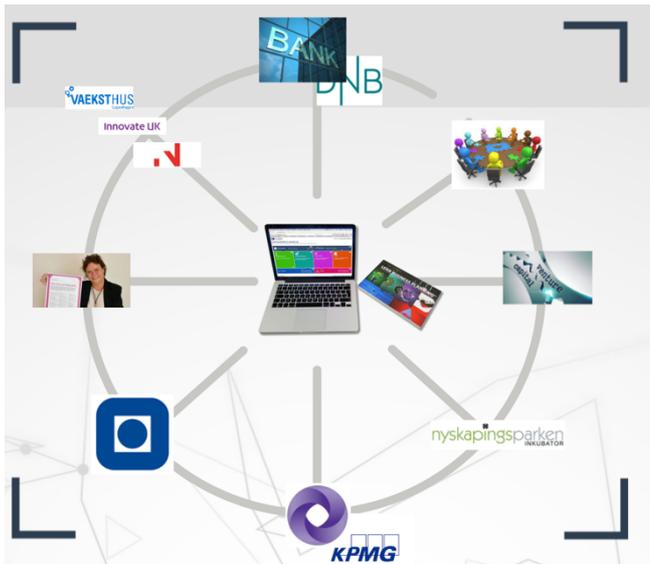

*Figure 3: The Entrepreneurial Ecosystem*

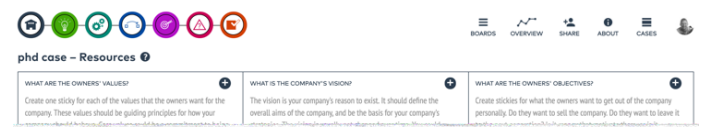

*Figure 4: Example of Card. One of the Business Idea Cards*

The design of the EP allows the Startups to document the changes as *Cards (as shown in* Figure 4*)*. As we will show later in the article, three different kinds of actions can be done to the Cards. The businesses can *Create* a new Card, *Update* a card and *Delete* an existing Card. Each of these four actions constitute an *Event*.

Different Case Companies may work with all of the Boards in different ways. They can gather all the stakeholders for a workshop, they can let everyone work independently, or they can give a smaller group the mandate to present a suggestion for discussions. One way or the other, they typically should end up agreeing on an initial set of Cards, before changing these to a larger or a smaller degree over time. There may also be differences within the Case from Board to Board. It may make sense to work with the strategic Boards (Resources, Business Idea, Business Model and Gaps) in a facilitated workshop including owners, board members and managers to different degrees. Then the departments may be given a mandate to suggest Objectives, before the management group consolidates them. Finally, they may let each key employee suggest his own Tasks before the department heads consolidate them.

Entrepreneurial information will be tracked by the system using seven boards: (1) Resource, (2) Business Idea, (3) Business Model, (4) Business Gap, (5) Objective, (6) Risk, and (7) Task.

The *Resource Board* have three boxes that are relevant for the EDW (as shown in Figure 5). The first give the different stakeholders a chance to define the *values* they want to base the business on. An organisational value is "*a belief that a specific mode of conduct is preferable to an opposite or contrary mode of conduct*" [15]. The next box fits Cards that suggest the *vision* of the business. A well-conceived vision consists of two major components: core ideology and envisioned future. The core ideology is unchanging while the envisioned future is what we aspire to become, to achieve, to create [4]. In the final box, the different owners of the Case Company can enter Cards with their personal *objectives* for the Case Company.

*Figure 5: The Resource Board*

The *Business Idea Board* (Figure 6) allows the stakeholders to define their *"Business Idea"*, or *"Mission"*. According to Canadian professor Chris Bart [1], the *Mission* consist of three components:
1. Key Contribution: What problem will you solve?
2. Key Market: Who will have this problem?
3. Distinction: What makes you unique?

The suggested work method will be to first populate the different Boxes with Cards in a brainstorming type process. Then the Companies are advised to link the different Cards to as many different Business Ideas as they please. Any Card can be used many times, and any Idea can include many Cards. The only rule is that there needs to be at least one Card in each Box. We expect significant variation in how many Business Ideas the different Companies will create, how many of them the will develop into Business Models, and how often they will Update them.

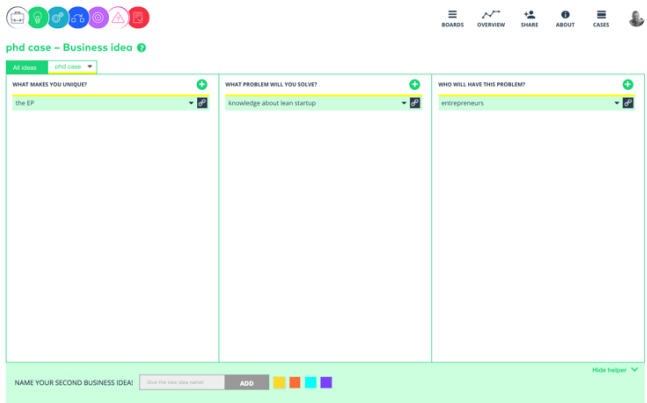

*Figure 6: The Business Idea Board*

The Business Model Board (Figure 7) allows the Case Companies to expand as many of their Business Ideas as they want to by adding answers to six questions: Who will be your early market customers? What will be your unique value proposition? What product features will you offer? What partners will you have? How will you sell? and How will you get paid? [6]. We will, as mentioned before, release a number of different templates of the EP. Some of these versions will use Osterwalder and Pigneur's BMC model [12] or Maurya's Lean Canvas [9] instead of our simplified Business Idea and Business Model. We will map these according to Appendix A, and flag which model has been used in each Case.

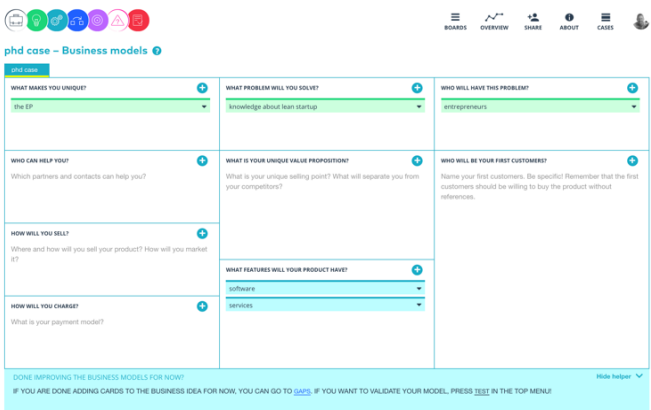

*Figure 7: The Business Model Board*

*The Gap Board* (Figure 8) is a combination of a competitor analysis and the Strength and Weaknesses part of a SWOT-analysis [8] The improvement from a traditional SWOT, is that all the strengths and weaknesses, both for the Case Company and their competitors, need to be derived from a specific Business Idea or Model element. In this way, we only discuss strengths and weaknesses that are relevant for the chosen Business Idea/Business Model.

In this part, the Case Companies will be asked to first define up to three competitors or groups of competitors. Then they will be led through all their active Business Idea and Model elements and asked to create strengths and weaknesses for as many as them as they see fit. Finally, they will be asked to do the same for each of their competitors.

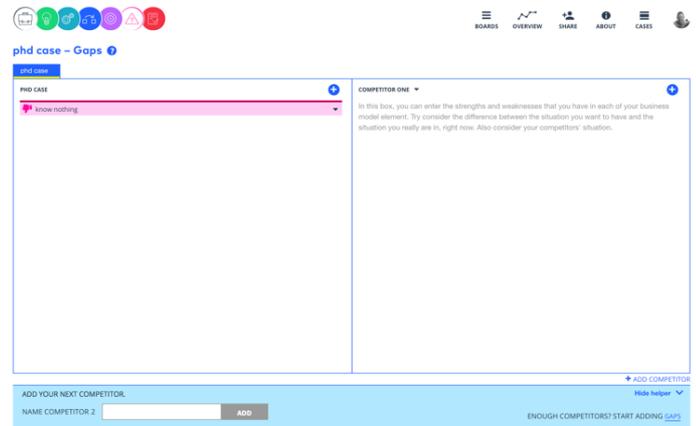

*Figure 8: The GAP Board*

The *Objectives Board* (Figure 9) is where the Case Company is supposed to actually start putting his strategy into action. The Case Company will be asked to start by defining their Skills Objectives. These are objectives that are related to increasing the competence of the organisation, typically by recruitment and hiring. These objectives can be defined as milestones or numerical objectives. Next the Case Company are asked to define their Product and Market Objectives. These Objectives are typically related to the development of new products or non-monetary KPI's regarding sales and marketing. These objectives can be defined as milestones or numerical objectives. Finally, the Case Company are asked to define their Money Objectives. These can be related to Revenues, Loans, Grants or Equity, and they are always stated in $ values. Money Objectives becomes the basis for the Revenue Part of the P&L Forecast.

As the Case period starts, the Case Company are asked to insert the actually achieved Objectives. They are also encouraged to adjust the estimates for the following periods as they gain experience. They are advised to delete irrelevant objectives and add new objectives at any point in the Case period.

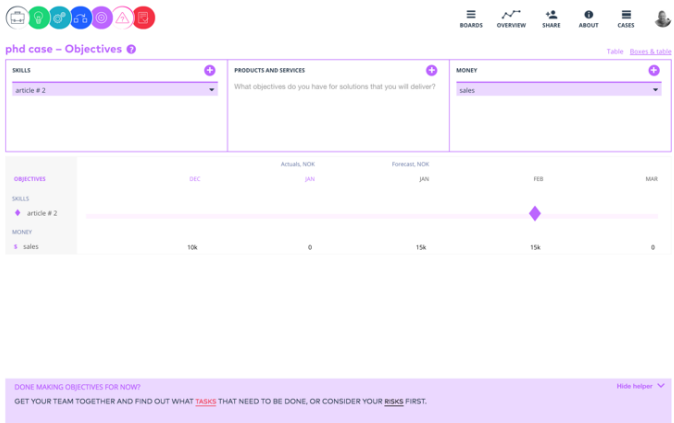

*Figure 9: The Objective Board*

The *Risk Board (*Figure 10*)* is where the Case Companies are advised to register their Opportunities and Threaths [8]. Like in the Gap Board, they are asked to go through their Business Idea and Business Model element one by one, and try to document Events related to them that may happen outside of the Case Company's control. Here they are asked to also go through their Objectives in the same way. They are asked to categorize each Risk (Opportunities and Threats) as low, medium or high probability and low, medium or high consequence.

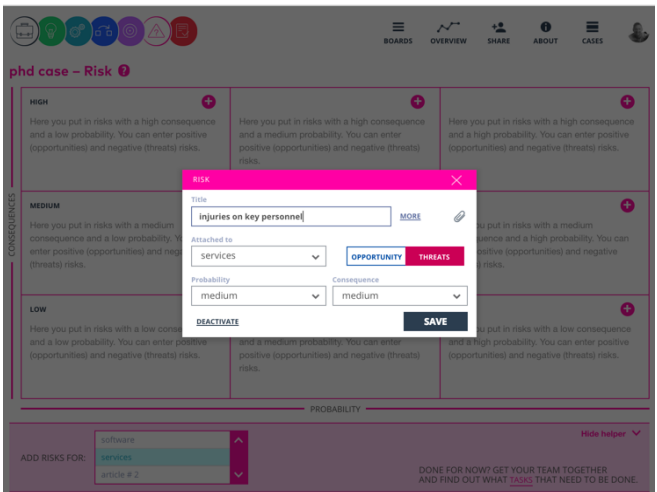

*Figure 10: The Risk Board*

The *Task Board (*Figure 11*)* is based on a KanBan model. The Case Companies are advised to go through each Objective and ask what must be done to reach each Objective. These Tasks are entered into the "Queue" Box. Here the level of inclusion of the team members may be very important. We suggest that each employee get a chance to define their own tasks, and then let department leaders make sure that no tasks are forgotten or done by several employees.

Each Task can be categorized as a Monthly Task or a one-off Task. Cost can be assigned to the Task, and the Task can be assigned to a specific employee. Tasks can then be moved from the Queue to the "Active" Boxes. In accordance to KanBan principles [10], it can always be moved back to the Queue. When done, the Task get moved over to the "Done" column, and the Actual Cost may be defined.

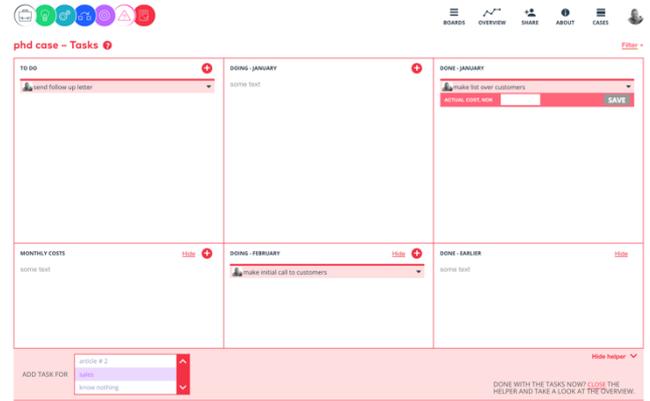

*Figure 11: The Task Board*

*Do we have a problem worth solving?* This is the initial customer test, as shown in Figure 12. We advise that the Case Company performs this test when they are working with the Business Idea. We also suggest that the Case Company should repeat the test every time they either do an iteration to the Business Idea, or they think there might be some kind of change in the preconditions for the Business Idea. The suggestion is to find 10-15 potential customers, and simply ask them to rate the problems that should be solved on a 7-point scale according to importance. The interviewees are encouraged to add their own suggestions for relevant problems, or to add comments to each of their ratings.

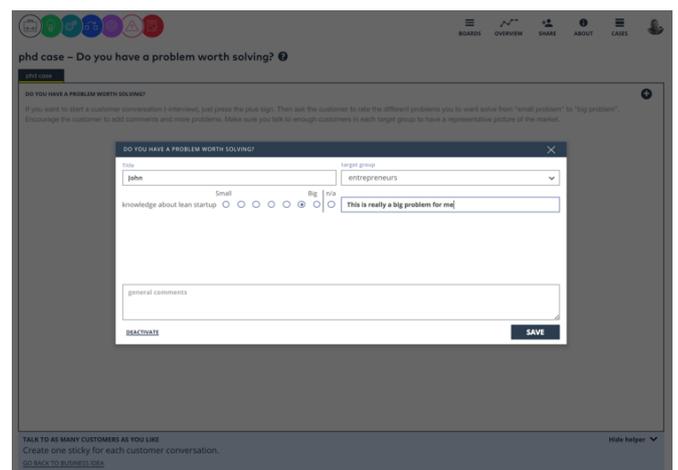

*Figure 12: Test one: Do we have a problem worth solving?*

The second customer test are: *Do we solve the problem? (* Figure 13). This should be done exactly the same way as the previous test, except it should be done while working on the business model, and that the interviewee is asked to rate the product features, sales methods and price models.

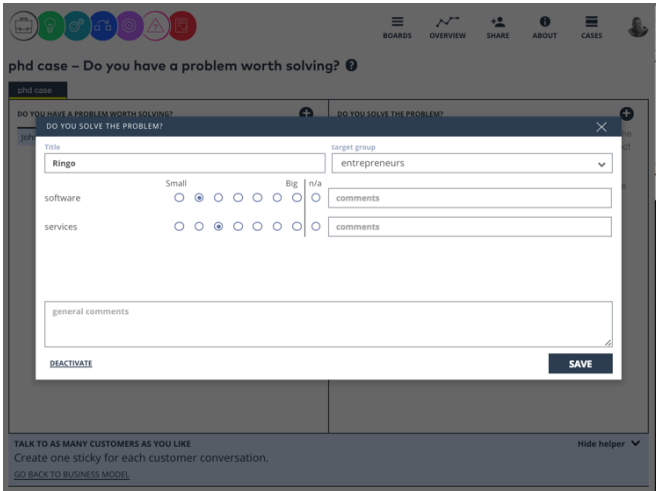

Figure 13: Test two: Do we solve the problem?

The last test we suggest is: *Is our market big enough?* (Figure 14). This test does not involve any kind of customer interviews. We ask the Case Company to make minimum and maximum estimates on: the total number of customers in each of their target markets, the expected market share they can expect and the expected monetary value of each of their customers. Based on these estimates, the system suggests a minimum and a maximum expected revenue for each market.

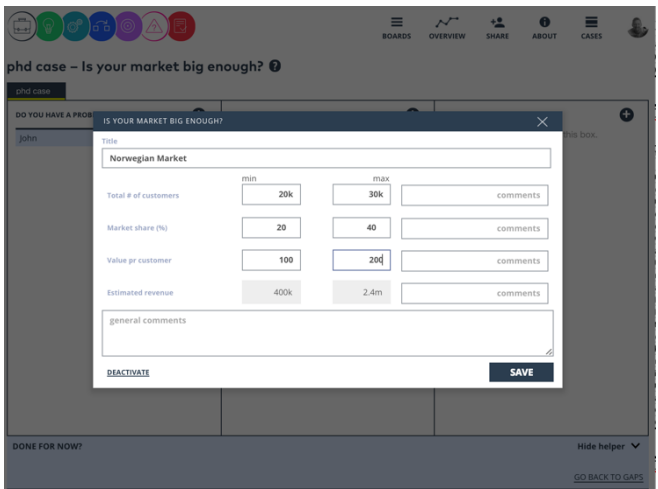

Figure 14: Test three: Is your market big enough?

The Overview Segment of the EP (as shown in Figure 15), is where the results of what is entered into the seven Boards are presented. Every month all unfinished Tasks and all Objectives in the period are presented in the End of Month part of the Overview. Here the Case Company are advised to do all necessary edits. They are also advised to go back and do alterations to their Business Idea, Business Model, Objectives or Tasks if the Financial Forecast looks critical.

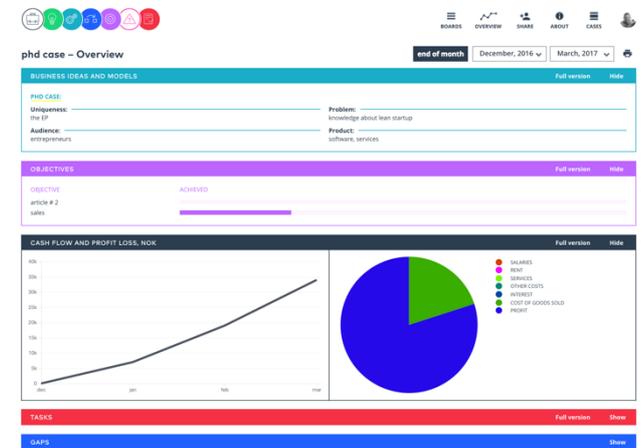

Figure 15: The Overview of a Case

*B. The PROFF-portal*

In Scandinavia, demographic criteria and financial performance indicators for different types of enterprises are public information in a central registration center. These data are gathered by a commercial company information portal called PROFF. We are building an interface toward their API to allow us to collect these data. We do this search on the organization number that the Case Company enters into the EP. Some of the data coming from PROFF, we label as Identifiers. That means that it is data connecting the Case Company to a name, a geographic area or an industry. Other data is financial success data such as revenue, profit etc. All these data is published annually, and we will have historical data for the last five years.

III. THE ENTREPRENEURSHIP DATA WAREHOUSE

In the EP, each Event is reflected in real-time. They are sitting there in a normalized form in a relational data structure meaning that it isn't possible to extract a record series quickly from it because many relations between different tables are to be joined. Doing this during research may severely affect the operation of the EP.

The data gathered in the EP will therefor be transferred to the EDW. The EDW is a non-relational database made in Elasticsearch that allows storage of big amount of data, specifically for research purpose. The EDW will also be periodically updated with Company information from the PROFF-portal. Our Extract/Transform/Load logic is being performed regularly. This procedure is pulling data from the EP, cleaning, denormalizing and pushing it into the EDW.

The EDW will be connected with a Kibana Visualization tool1 from the start. Eventually, this data will be made available for a large spectre of analytics tools.

*Table 1: Event categories*

| # | Source | Event Category |
|---|--------|----------------|
| 1 | ADM | Participants |
| 2 | ADM | Case Settings |
| 3 | RES | Values |
| 4 | RES | Vision |
| 5 | RES | Owner's Objectives |
| 6 | BI | Business Idea |
| 7 | BI | Key Contribution |
| 8 | BI | Key Market |
| 9 | BI | Distinction |
| 10 | BM | Early Market Customer |
| 11 | BM | Unique Value Proposition |
| 12 | BM | Product Feature |
| 13 | BM | Partner |
| 14 | BM | How to Sell |
| 15 | BM | How to Get Paid |
| 16 | GAP | Strength & Weaknesses |
| 17 | OBJ | Objectives |
| 18 | RISK | Opportunities & Threats |
| 19 | TASK | Task |
| 20 | TEST | Problem Worth Solving? |
| 21 | TEST | Solve the Problem? |
| 22 | TEST | Market Big Enough? |
| 23 | PROFF | Registration |
| 24 | PROFF | Revenue |
| 25 | PROFF | Profit & Loss |
| 26 | PROFF | Balance Sum |
| 27 | PROFF | Return On Assets |
| 28 | PROFF | Profit & Loss Percentage |
| 29 | PROFF | Return on Equity |
| 30 | PROFF | Current Ratio |
| 31 | PROFF | Equity Ratio |
| 32 | PROFF | Gearing |
| 33 | PROFF | Registration or Bankruptcy |
| 34 | PROFF | Number of Employees |

The EDW will be built as a single table structure containing time series Events gathered from either the EP or PROFF. Each Event will have a unique time stamp and contain a number of identifying data.

The unique opportunity offered by our setup is that every Event belongs to one out of 34 *Event Categories*, as shown in Table 1.

Since all Events in the same Category is made in a very structured manner, they will have the same basic meaning – regardless of who initiated the Event, what Company they work in, or even what language they are using. This uniformity of the Event Categories is what allows us to compare the Events between Cases, countries, lifecycle phases and industries.

---

1 https://www.elastic.co/products/kibana

The relation between the Events, the Event Categories and time is illustrated in Figure 16. There can be an unlimited number of Events within each Event Category, and there will be one such combination for each Case. Each Company in the system can have several Cases.

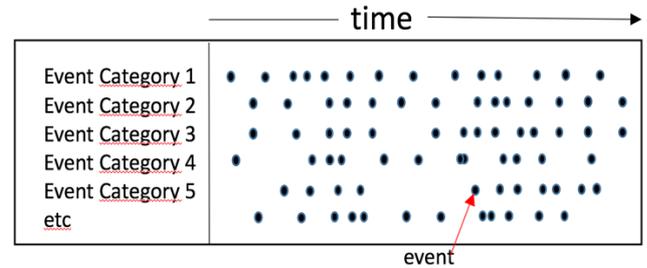

*Figure 16: Events over time*

Each event will consist of a number of Datafields. The full structure of the Datafields are presented in Table 2. First we will introduce the Datafields that are included in Events from all Event Categories:

*Table 2: Datafields included in all Event Categories*

| Case ID |
|---|
| Case title |
| Event ID |
| Timestamp |
| Event category |
| Action type (CUD) |
| Case Participant |
| Company Name |
| Organization Number |
| Country |
| Postcode |
| NACE-code (Industry) |
| Added by Case Role |
| Client ID |
| Relating to whole company |
| Event Title |
| Event Description |

Since we are organizing the DW as a single table data structure, there will be a set of data that will be included in all the Events.

Each Event will be linked to a Case via a unique Case ID number and a Case title. It will also be linked to a specific Business Idea/Model via an Idea/Model title. Each Event will also be given an Event ID number, a Timestamp, a link to the relevant Event Category and an Action Type to tell whether the Event was created, updated or deleted. The individual initiating the Event will be linked to the event via a Case Participant ID. We will link each Event to a Company by registering the Company name, Organization number and Country. Based on the Organization, we will verify the Company name and country in the PROFF Portal. Finally, we will gather the postcode and the NACE-code from the PROFF portal. The NACE-code is a unique nomenclature defining which industry the Company is in.

The Case Role defines whether the Event has been initiated by an Entrepreneur, an Enabler or an Educator and the Client ID defines what template the Event is initiated from.

*Table 3: Event Category Specific Fields*

| Source | Field Name | Event Category |
|---|---|---|
| Res | Case Participant Type | Participants |
| Adm | Business Model Type | Case Settings |
|  | Timespan | Case Settings |
|  | Forecast Type | Case Settings |
| Gap | Strength & Weaknesses | Strength & Weaknesses |
|  | Competitor Name | Strength & Weaknesses |
| Object | Objective Category | Objectives |
|  | Objective Type | Objectives |
|  | Actual VS Forecast | Objectives |
|  | Objective Month | Objectives |
|  | Value | Objectives |
| Risk | Opportunities & Threats | Opportunities & Threats |
|  | Probability | Opportunities & Threats |
|  | Consequence | Opportunities & Threats |
| Task | Cost Group | Task |
|  | Task Month | Task |
|  | Actual VS Forecast | Task |
|  | Value | Task |
|  | Status | Task |
| Test | Average Score | Problem Worth Solving? |
|  | Average Score | Solve the Problem? |
|  | Total # of customers low | Market Big Enough? |
|  | Total # of customers high | Market Big Enough? |
|  | Market share low | Market Big Enough? |
|  | Market share high | Market Big Enough? |
|  | Value per customer low | Market Big Enough? |
|  | Value per customer high | Market Big Enough? |
| Proff | Yearly Revenue | Revenue |
|  | Yearly Profit and Loss | Profit & Loss |
|  | Yearly Balance Sum | Balance Sum |
|  | Yearly Return on Assets | Return On Assets |
|  | Yearly Profit and Loss % | Profit & Loss Percentage |
|  | Yearly Return on Equity | Return on Equity |
|  | Yearly Current Ratio | Current Ratio |
|  | Yearly Equity Ratio | Equity Ratio |
|  | Yearly Gearing | Gearing |
|  | Number of Employees | Number of Employees |

The EP will be delivered in a wide range of different templates, each containing chosen Boards. Different templates will be published by different partnering organizations. Virgin are releasing one version to manage UK Startup Loans, Grant Thornton are releasing a different one to do business development in the Swedish market whilst Innovation Norway are releasing one specific one for cultural entrepreneurs in Norway. We will register which of the templates any given Case is utilizing. It is also defined whether the case was Relating to whole company or not.

In addition to these fields, each event will contain a title and an extended information field. It may also contain one or more category specific fields.

The Resource Category specific field is whether a Case Participant is a partner or an employee.

For the Admin Board Categories, we get three specific fields. The Companies can also choose between three different business model structures. Osterwalder and Pigneur's "BMC" [12], Maurya "Lean Canvas" [9] and our own "Lean Business Canvas" [5]. We will register what canvas method each Company are choosing. We will also register whether the plan length is 3,6,9 or 12 months. Next, we need to know if a rolling forecast is chosen or not. A rolling forecast means that a new month is added to the end of the Case period at the beginning of every month.

From the Gap Board, we have a Category Specific Field to categorize each Gap as a strength or a weakness and one to define the name of the company the strength or weakness is relating to. This may be the Case Company or any of the three defined Competitors.

For the Objectives Board, we have five specific fields. The Objective Categories defines whether the Event is a Skills, Product/Market or a Money Objective. The Objective Type whether it is a milestone, numerical, revenue, loan, equity or grant objective. We define if we have a forecasted or an actual Objective and what Month it relates. If we have a numerical or monetary Objective, the value is defined.

We have opportunities and threats from the Risk Board. Here we store the estimated probability (High/ medium/ low) and consequence (High/ medium / low) of each risk.

From the Task Board, we also have five specific fields. We store the cost group and value on the Tasks that have that, and the month the task relates to. In addition to the regular registrations of creation, edits and deletions – we also register each time the Task has been moved between the queue and active status and if we have a forecasted or an actual Task.

The ten Specific Fields coming from the Test Boards starts with whether we have any problems or features added by the Customer in the interview. For the "do they have a problem worth solving?" and "do they solve the problem?" we store the average scores from each interviewee. For the "is the market big enough?" Test, we store the high and low estimates of the total number of customers, market share percentage and value per customer.

For the Swedish, Danish and Norwegian firms where we have an organization number, we will be able to get financial and demographic data from the PROFF portal. The last ten Specific Fields relate to the Event Categories originating from these data. We will be able to get five years of financial data. The particular Events we will store in the DW will be; revenue, profit/loss, balance sum, return on assets, profit and loss percentage, return on equity percentage, current ratio percentage, equity ratio and gearing percentage. We will register any change in the number of employees. We will try to gather as much of these data as possible also for Companies coming from outside the Nordic countries. However, the degree of openness with regard to these figures for startups are especially high in Scandinavia.

## IV. THREATS TO VALIDITY

We are building an information gathering setup that is designed to gather a large amount of time series information from tens of thousands of Companies all over the world. There will be some methodological challenges to this operation. In this chapter, we will try to describe them:

### A. Actions outside of the system

The main problem with our approach is that we can not be sure that all the activities of the Case Companies will be documented in the EP. There is of course a possibility that both changes to the Business Idea, Business Model, Project Development and Customer Interaction will be performed without anybody registering it in the system. We will try to counteract this by doing test interviews with selected Case Companies.

### B. The self-selection bias

In accordance with privacy laws, all Companies entering Cases into the EP will have the option to opt out of volunteering its data for research purposes. It is imaginable that a certain subset of participants will have a larger tendency to do so than other. As an example, Companies in sectors with intense competition might be more eager to keep it´s data private than others.

We think it will be possible to manage this problem by doing qualitative Case studies of some of the most interesting Cases in the system – and in those, focus on this specific question.

### C. The Company/Case problem

The level of analysis in the EP is *Case*. What is a Case is defined by each Case Owner. There is no 1:1 relationship between a Case and a Company. Some of the Cases registered will not relate to a Company at all, they will merely be entrepreneurial ideas in the pre-establishment phase. These Cases will not be linked to a Company – and will as such not influence the study.

Another scenario is that a Case is just reflecting to a specific sub-process within a Company, such as a spin-off, a department or a project. We will handle this by asking specifically whether the Case is relating to the whole Company or not.

### D. The change of Identifiers problem

Some of the Identifiers may change over time. We can imagine that the company may change it's name or moves from one geography to another. We will simply register such changes. The two unchanging Identifiers will be the Case number generated from the system and the Organization number.

### E. The time atomization challenge

The core of our system is the time series Events that are gathered from the EP and PROFF. There are two problems with the time atomization of the time stamps.

First, we have to make some sort of estimate of what constitutes an Event. Say that you are Creating an Event. You open the Event Card, enter your content and save it. Then you remember that you have forgotten to input something, and open the card 10 seconds later – and enter something more. This sequence will be registered as two Events. One Create Event, and one Update Event. For now, we will keep it like that, but depending on what we learn from the data, we might put a time limit on when such an edit shall be deemed as an individual Update.

Second, there are some Events that takes place over a time range. As an example, any yearly financial update from PROFF will relate to a whole calendar year. We will solve this by assigning such an Event to the last second of the time range. So, the revenue of the calendar year of 2017 will be assigned to the time stamp 31.12.2017 at 23.59.59.

## V. POSSIBLE RESEARCH METHODS AND INITIAL DATA TEST

This article focuses on the design of the Entrepreneurship Platform, The Entrepreneurship Data Warehouse and the general architecture to gather a large amount of behavioural, demographic and financial time series data from as many and as diverse entrepreneurial endeavours as possible. Having these data will allow us to perform any kind of analysis, and the specific analysis design will be the theme for future articles. However – our present belief is that we do our analysis in three steps:

Step 1: we want to do an initial hypothesis finding by doing an explorative quantitative analysis on a segment of the data set. We will start doing time series analysis using principal component analysis [7] from the launch of the final version of the test harness in May 2017.

Step 2: Based on the hypothesis we find in Step 1, we will do quantitative hypothesis verification/ falsification on another segment of the data set.

Step 3: finally, we will do depth analysis of the most interesting Cases, including qualitative analysis of the content of each element, studies of the financial data from each Case and interviews of the key personnel in the Case Companies. In this phase we can also do verification tests with Case Companies from areas where we do not have external financial data. We will utilize automated word and text analysis in this phase of the study.

In our previous article [6] we did a brief test to see whether entrepreneurs actually behaved in a lean way. In this stage, we want to verify that we are able to build up a data warehouse with an acceptable growth in users to get the kind of big data analysis that we are searching for. Therefore we want to use Kibana to check that our user base grows in an acceptable rate, and then contains the data that we need it to contain. The first full period of operations for the entrepreneurship platform 1.0 was February $1^{st}$ – May 15th 2017. During this period, we had 78.296 Events distributed on 1.377 Cases in the system. We have transferred a small subset of time series Events into the EDW for this period. The selected Datafields are:

*Table 4: Initial Data Fields*

| 1. Case title | 2. Event category |
|---|---|
| 3. Event title | 4. Action type (CUD) |
| 5. Timestamp | 6. Card Title |

So we had an average of almost 57 Events per Case in this period. The Events where distributed over the following 22 Event Categories (We have not integrated the PROFF Event Categories yet):

*Table 5: Events per Event Category*

|  | Create | Update | Delete |
|---|---|---|---|
| **Participants** | 2688 | 64 | 783 |
| **Case Settings** | 1194 | 8565 | 49 |
| **Values** | 1486 | 458 | 180 |
| **Vision** | 804 | 334 | 99 |
| **Owners Objectives** | 926 | 298 | 95 |
| **Business Idea** | 1317 | 2106 | 239 |
| **Key Contribution** | 3356 | 1922 | 542 |
| **Key Market** | 2754 | 818 | 357 |
| **Distinction** | 3266 | 1385 | 549 |
| **Early Market Customer** | 1478 | 376 | 164 |
| **Unique Value Proposition** | 1316 | 517 | 200 |
| **Product Feature** | 3650 | 690 | 1021 |
| **Partner** | 3297 | 740 | 790 |
| **How to Sell** | 1696 | 324 | 177 |
| **How to Get Paid** | 1091 | 229 | 125 |
| **Strength & Weaknesses** | 3367 | 1119 | 286 |
| **Objectives** | 2568 | 3275 | 577 |
| **Opportunities & Threats** | 929 | 364 | 65 |
| **Task** | 4324 | 5421 | 748 |
| **Problem Worth Solving** | 354 | 135 | 69 |
| **Solve the Problem** | 67 | 40 | 9 |
| **Market Big Enough** | 40 | 22 | 2 |

The data will of course be more interesting when we are able to monitor them over time. Even further so when we will be able to compare them to the PROFF Event Categories.

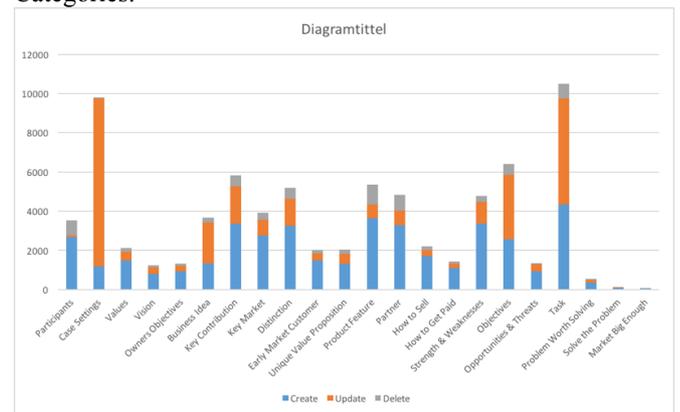

*Figure 17: Event per Event Category illustration*

However, we can already see some indications. Given that the Entrepreneurship Platform has so far not been marketed at all, we can see that there is a good chance that we will be able to get a significant amount of Data. Further more we see that we get a good number of Events in each Event Category. We get the most Events in the Task Category and the least in the Test Categories. This is not surprising. We know that actually going out and performing customer interviews is quite a challenge for many Entrepreneurs. We also have a relative low number of Events in the Values, Vision and Owner's Objective Categories, whereas both the Business Idea and Business Model Categories seems to have more Events. This may indicate that the initial entrepreneurs distribute their attention evenly between Business Ideas, Business Model, Objectives and Tasks – and use less time on (the less practical) Values, Vision and Owner's Objective Categories.

The distribution between the Create, Update and Delete Action Types are quite similar in the different Event Categories, apart from Case Settings. This Category have a much higher percentage of Updates. The explanation for this will have to wait until we get the full data set in the Data Warehouse.

Next, we have sorted the Events over the three and a half months we have been in operation. Given that we only have data from half of May, and that most of April is Easter – we see a steady flow of Events over the months.

When we include the Event Categories in the analysis, we see the same picture. The Event distribution throughout the Event Categories does generally not vary from month to month. The only anomaly here is the higher number of Case Settings Events in April. To find an explanation for this, we will need to get the full dataset into the Data Warehouse.

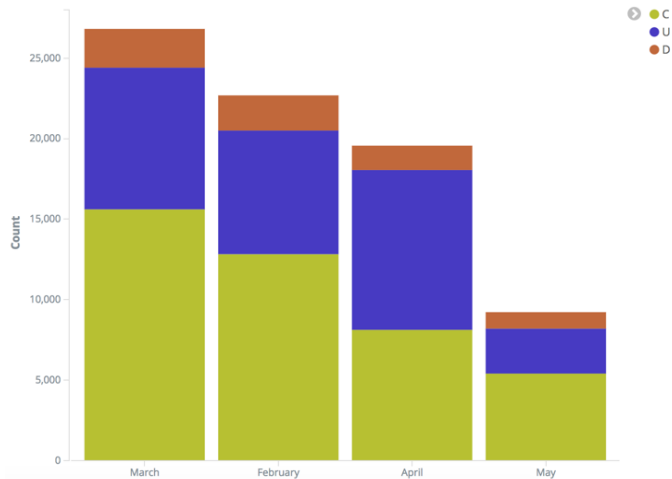

*Figure 18: Event per month & Action Type illustration*

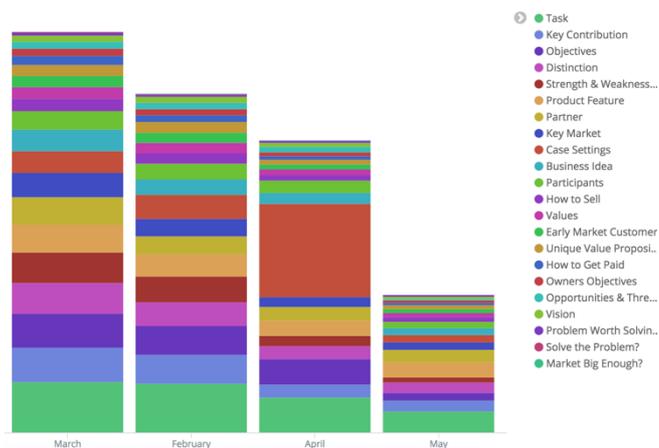

*Figure 19: Event per month & Event Category illustration*

## VI. CONCLUSIONS

In our last paper [6], we concluded that it will be possible to find Entrepreneurs that have a "Lean" behaviour, meaning that they will incrementally develop their business over time. In this paper we can conclude that it is possible to motivate Entrepreneurs to leave time series data regarding their business development process by giving them access to a business development tool, and that those data can be effectively transferred to a Data Warehouse.

## APPENDIX A

*Table 6: Comparing our model and others*

| Our Model | BMC | Maurya |
|---|---|---|
| key contribution | no match | problem |
| key market | customer segments | customer segments |
| distinction | no match | unfair advantage |
| early market customers | no match | no match |
| unique value proposition | value proposition | unique value proposition |
| product features | no match | solution |
| partners | key partners | no match |
| how the Startups sell | channels | channels |
| how the Startups get paid | revenue streams | revenue streams |
| no match | no match | key metrics |
| no match | key activities | no match |
| no match | relationships | no match |